\documentclass[journal]{IEEEtran}
\usepackage{cite}

\ifCLASSINFOpdf
  \usepackage[pdftex]{graphicx}
  \graphicspath{{../figures/}}
  \DeclareGraphicsExtensions{.pdf,.jpeg,.png}
\else
\fi
\usepackage{amsmath}
\usepackage{amssymb}
\usepackage{amsfonts}
\usepackage{amsthm}

\usepackage{url}
\usepackage{xcolor}

\usepackage{float}
\usepackage{textcomp}
\floatstyle{ruled}
\newfloat{algorithm}{tbp}{loa}
\providecommand{\algorithmname}{Algorithm}
\floatname{algorithm}{\protect\algorithmname}

\usepackage{nicefrac}\usepackage{chngcntr}

\providecommand{\U}[1]{\protect\rule{.1in}{.1in}}
\counterwithin{figure}{section}

\newtheorem{definition}{Definition}\newtheorem{problem}{Problem}

\makeatother

\begin{document}

\title{Developments in Mathematical Algorithms and Computational Tools for
Proton CT and Particle Therapy Treatment Planning}

\author{Yair~Censor, Keith~E.~Schubert,~\IEEEmembership{Senior~Member,~IEEE,} and~Reinhard~W.~Schulte,~\IEEEmembership{Member,~IEEE}. \newline Preprint accepted for publication in IEEE Transactions on Radiation and Plasma Medical Sciences. \newline August 16, 2021.
\thanks{Y. Censor is with the Department of Mathematics, The University of
Haifa Haifa, Israel}\thanks{K. E. Schubert is with the Department of Electrical and Computer Engineering,
Baylor University, Waco, TX, USA}\thanks{R. W. Schulte is with the Department of Basic Sciences, Loma Linda
University Loma Linda, CA, USA}}
\ifCLASSOPTIONpeerreview


\fi
\maketitle

\begin{abstract}
We summarize recent results and ongoing activities in mathematical
algorithms and computer science methods related to proton computed
tomography (pCT) and intensity-modulated particle therapy (IMPT) treatment
planning. Proton therapy necessitates a high level of delivery accuracy
to exploit the selective targeting imparted by the Bragg peak. For
this purpose, pCT utilizes the proton beam itself to create images.
The technique works by sending a low-intensity beam of protons through
the patient and measuring the position, direction, and energy loss
of each exiting proton. The pCT technique allows reconstruction of
the volumetric distribution of the relative stopping power (RSP) of
the patient tissues for use in treatment planning and pre-treatment
range verification. We have investigated new ways to make the reconstruction
both efficient and accurate. Better accuracy of RSP also enables more
robust inverse approaches to IMPT. For IMPT, we developed a framework
for performing intensity-modulation of the proton pencil beams. We
expect that these developments will lead to additional project work
in the years to come, which requires a regular exchange between experts
in the fields of mathematics, computer science, and medical physics.
We have initiated such an exchange by organizing annual workshops
on pCT and IMPT algorithm and technology developments. This report
is, admittedly, tilted toward our interdisciplinary work and methods.
We offer a comprehensive overview of results, problems, and challenges
in pCT and IMPT with the aim of making other scientists wanting to
tackle such issues and to strengthen their interdisciplinary collaboration
by bringing together cutting-edge know-how from medicine, computer
science, physics, and mathematics to bear on medical physics problems
at hand.
\end{abstract}




\IEEEpeerreviewmaketitle{}

\section{Introduction}

\label{sec:aims} Proton therapy is becoming increasingly common for
cancer radiation therapy. Protons afford tissue-sparing advantages
that should be carefully tested against the best available photon
therapy techniques, i.e., intensity-modulated radiation therapy (IMRT).
However, this requires further development of image-guidance and inverse
planning techniques for proton and ion therapy. This report summarizes
some developments of advanced algorithmic and computational methods
for proton computed tomography (pCT) and for intensity-modulated particle
therapy (IMPT) that we developed by applying expertise in optimization
algorithms, computer science, and medical physics, related to proton
and ion therapy. We give references to published work when applicable
and include details about work-in-progress. The purpose of over-viewing
our work in this rapidly evolving field is to inspire other researchers
to contribute to it in an interdisciplinary fashion.

Both IMPT and pCT are ``inverse problems''; they are processes of
calculating causal factors from a set of observations they produced.
For example, in IMPT or IMRT the fluence of individual pencil beams
or beamlets will produce a dose distribution and the clinical prescriptions
play the role of ``observations'' in this framework from which the
proper fluences are calculated. Likewise, in pCT, relative stopping
power (RSP) is the property of tissues that results in energy loss
of protons and ions. Here, measured energy loss of individual protons
traversing the object are the ``observations'' and the RSP is calculated
from these observations, forming the pCT image. Starting from this
premise, we discuss the following topics mostly from the mathematical
and computational points of view:
\begin{itemize}
\item The development of a feasibility-seeking approach and algorithms for
treatment planning of IMPT.
\item The superiorization methodology (SM) combined with the diagonally
relaxed orthogonal projection (DROP) feasibility-seeking algorithm.
\item The application of the SM to total variation superiorization (TVS)
for pCT reconstruction.
\item The use of blob basis functions and the reconstruction from pCT data
acquired in the presence of organ motion.
\item The derivative-free framework for the SM with component-wise perturbations.
\item The selection of appropriate digital anthropomorphic phantoms for
exploring pCT and IMPT.
\item The selection of simulation tools to generate pCT data and, as a forward
calculation method, to calculate dose contributions from proton pencil
beams.
\item The development of tools to compare pCT and IMPT results with those
of concurrent methods.
\item The development of code optimized for general purpose graphics processing
units (GPGPU) that combines algorithmic advances, memory, and computational
tuning, required to achieve speedup.
\item The efforts to acquire pCT data with phantoms using the preclinical
pCT scanner at the Northwestern Medicine Chicago Proton Center (NMCPC).
\end{itemize}

\section{Model Formulations and Mathematical Algorithms for pCT and IMPT Inverse
Problems}

\label{sec:methods} In this section, we give an overview of some
of the methods used in the investigations of the topics listed above.
If available, we point to additional details about the methods in
the published literature. From the mathematical modeling and algorithmic
points of view, we take recent algorithmic developments and ``translate''
them to applications in pCT and IMPT inverse problems. We demonstrate
this approach by two specific novel techniques that were investigated
and have been published but can benefit from further adaptation and
refinement for these applications. These are the \textbf{superiorization
methodology} for pCT image reconstruction and the \textbf{split inverse
problem paradigm} for IMPT.

The core of the fully-discretized inverse problems of IMPT and pCT
problems discussed here consists of the constraints, which are dictated
by the underlying physical model. In the \textit{feasibility approach},
we look at \textit{convex feasibility problems }(CFPs) of the form:
find a vector $x^{\ast}\in C:=\cap_{i=1}^{I}C_{i},$ where the sets
$C_{i}\subseteq R^{J}$ are nonempty closed convex subsets of the
Euclidean space $R^{J}$, see, e.g., \cite{bb96,byrnebook,chinneck-book}
or \cite[Chapter 5]{CZ97} for results and references on this broad
topic.

Both IMPT inverse planning and pCT, as well as other inverse problems
where the underlying system has very large values of $I$ and $J$
and is often very sparse, fall under this category. Under these circumstances,
\textit{projection methods} are beneficial \cite{bb96,bcgh,bc01,cdz04,cms07,Bauschke-Koch}.
These are iterative algorithms that use projections onto sets, e.g.,
hyperplanes or half-spaces in the linear cases, while relying on the
general principle that when a family of, usually closed and convex,
sets is present, projections onto the given individual sets are easier
to perform than projections onto other sets (intersections, image
sets under some transformation, etc.) that are derived from the given
individual sets.\footnote{The fully-discretized formulation of the inverse problem of IMRT treatment
planning was not commonly accepted in the early days as it is today.
In 1982, Brahme, Roos, and Lax tried to solve the inverse problem
of radiation therapy treatment planning in its continuous (not fully-discretized)
formulation via integral inversion \cite{Brahme1982}. However, to
be able to generate the required inverse transform, they had to make
unrealistic assumptions on the model that rendered their analysis
impractical. Altschuler and Censor proposed in 1984 \cite{Alt-Cen1984}
a fully-discretized IMRT model. Here, ``fully-discretized'' means
that not only the irradiated volume is discretized into voxels but
that also the external radiation field is discretized into ``rays''
(``beamlets'' or ``pencil beams'' in today's language). The initial
conference report was followed by a sequence of papers that established
this approach \cite{CAP-AMC1988,CAP-IP1988,CPA1987}. To the best
of our knowledge, these and the independent 1990 paper of Bortfeld
et al. \cite{bortfeld-1990} were the first publications that suggested
the fully-discretized approach to the inverse problem of IMRT, see
also \cite{Censor1999} and the introduction of \cite{cho-marks-2000}.}

\subsection{The Superiorization Methodology}

\label{sec:SM} The superiorization methodology (SM) can be applied
to the data of constrained minimization (CM) problems of the form:
$\mathrm{minimize}\left\{ \phi(x)\mid x\in C\right\} ,$ where $\phi:R^{J}\rightarrow R$
is a target function, and $C\subseteq\Theta\subseteq{R^{J}}$ is a
given feasible set defined by constraints, see, e.g., \cite{hgdc12}.
It aims at finding a feasible point that is superior (with respect
to the target function value) to one returned by an algorithm that
is only feasibility-seeking. In doing so, SM is situated between feasibility-seeking
and full-fledged CM.

There are two main reasons why superiorization is beneficial: (i)
for a problem for which an exact CM algorithm has not yet been discovered,
but there are iterative feasibility-seeking methods that provide constraints-compatible
solutions, which can be turned by the superiorization methodology
into methods that will be practically useful for the target function
reduction effort; and (ii) when existing exact optimization algorithms
are either very time consuming or require too much computer space
for large problems to be processed by run-of-the-mill computers. On
the other hand, space- and time-efficient algorithms exist for constraints-compatibility-seeking,
and these can be turned into efficient algorithms for superiorization.
Examples of such situations are given in \cite{compare13,DHC09}.

There is no general answer yet to the question under
which circumstances one should resort to superiorization as the method
of choice. Even when tractable constrained optimization algorithms
are available, the SM often yields better or comparable results. This
has been demonstrated in \cite{compare13} where a comparison between
the projected subgradient method and SM showed that the SM performed
better when applying it to an image reconstruction from projections
problem. Many works cited on \cite{sup-bib} attest to the practical
success of the SM in a variety of situations.

In the SM, one associates with the feasible set $C$ a proximity function
${Prox}_{C}:\Theta\rightarrow R_{+}$, which is an indicator of how
incompatible a vector $x\in\Theta$ is with the constraints. For any
given $\varepsilon>0$, a point $x\in\Theta$ for which ${Prox}_{C}(x)\leq\varepsilon$
is called an $\varepsilon$\textit{-compatible point} for $C$.

The $\varepsilon$-output of a sequence, defined in \cite{CDH10},
can be explained as follows. For a nonnegative $\varepsilon$ and
a sequence $R:=\left(x^{k}\right)_{k=0}^{\infty}$ of points in $\Theta$,
the $\varepsilon$-output of the sequence $R$ is a point $x\in\Theta$
that has the following properties: ${Prox}_{C}(x)\leq\varepsilon,$\emph{
}and there is a nonnegative integer $K$ such that $x^{K}=x$ and,
for all nonnegative integers $k<K$, we have ${Prox}_{C}(x^{k})>\varepsilon$\emph{
}. If there is such an $x$, then it is unique. If there is no such
$x$, then we say that the $\varepsilon$-output of the sequence is
\textit{undefined} and otherwise, that it is \textit{defined}.

In order to ``superiorize'' a feasibility-seeking algorithm, commonly
called the ``Basic Algorithm'', represented by the iteration $x^{k+1}=\boldsymbol{A}_{C}(x^{k}),$
we need it to have \textit{strong perturbation resilience }in the
sense that for every $\varepsilon>0,$ for which an $\varepsilon$-output
is defined for a sequence generated by the Basic Algorithm for every
$x^{0}\in\Theta$, we have also that the $\varepsilon^{\prime}$-output
is defined for every $\varepsilon^{\prime}>\varepsilon$ and for every
sequence $\left\{ y^{k}\right\} _{k=0}^{\infty}$ generated by
\begin{equation}
y^{0}=x^{0},\:\:y^{k+1}=\boldsymbol{A}_{C}\left(y^{k}+\beta_{k}v^{k}\right),\text{ for all }k\geq0,\label{eq:perturb}
\end{equation}
where the vector sequence $\left\{ v^{k}\right\} _{k=0}^{\infty}$
is bounded and the scalars $\left\{ \beta_{k}\right\} _{k=0}^{\infty}$
are such that $\beta_{k}\geq0$, for all $k\geq0,$ and $\sum_{k=0}^{\infty}\beta_{k}<+\infty$.
See, e.g., \cite{compare13,CDH10} for additional details.

Sufficient conditions for strong perturbation resilience of a Basic
Algorithm were derived in \cite[Theorem 1]{hgdc12} and \cite{compare13}.
Along with the constraints $C\subseteq R^{J}$, we look at a target
function $\phi:\Theta\subseteq R^{J}\rightarrow R$, with the convention
that a point in $R^{J}$ for which the value of $\phi$ is smaller
is considered \textit{superior} to a point in $R^{J}$ for which the
value of $\phi$ is larger. The essential idea of the superiorization
methodology is to make use of the perturbations of (\ref{eq:perturb})
to transform a strongly perturbation resilient algorithm that seeks
feasibility into an algorithm with an output that is equally good
for constraints-compatibility but is also superior (not necessarily
optimal) according to the target function $\phi$.

The SM accomplishes its goal by producing from the Basic Algorithm
another algorithm, called its \textit{superiorized} version, that
implements, at every iteration, a perturbation that reduces the target
function value locally, i.e., $\phi\left(y^{k}+\beta_{k}v^{k}\right)\leq\phi\left(y^{k}\right)$.
The Superiorized Version of the Basic Algorithm assumes that we have
available a summable sequence $\left\{ \eta_{\ell}\right\} _{\ell=0}^{\infty}$
of positive real numbers (for example, $\eta_{\ell}=a^{\ell}$, where
$0<a<1$) and it generates, simultaneously with the sequence $\left\{ y^{k}\right\} _{k=0}^{\infty}$
in $\Theta$, sequences $\left\{ v^{k}\right\} _{k=0}^{\infty}$ and
$\left\{ \beta_{k}\right\} _{k=0}^{\infty}$. The latter is generated
as a subsequence of $\left\{ \eta_{\ell}\right\} _{\ell=0}^{\infty}$,
resulting in a nonnegative summable sequence $\left\{ \beta_{k}\right\} _{k=0}^{\infty}$.
The algorithm further depends on a specified initial point $y^{0}\in\Theta$
and a positive integer $N$. It makes use of a logical variable called
\textit{loop}\emph{. }Such a superiorized algorithm is presented here
by its pseudo-code in Algorithm 1.\medskip{}

\begin{algorithm}
\textbf{Algorithm 1: The Superiorized Version of the Basic Algorithm}

\textbf{set} $k=0$

\textbf{set} $y^{k}=y^{0}$

\textbf{set} $\ell=-1$

\textbf{repeat}

$\qquad$\textbf{set} $n=0$

$\qquad$\textbf{set} $y^{k,n}=y^{k}$

$\qquad$\textbf{while }$n$\textbf{$<$}$N$

$\qquad\qquad$\textbf{set }$v^{k,n}$\textbf{ }to be a non-ascending
vector for $\phi$ at $y^{k,n}$

$\qquad$\textbf{$\qquad$set} \emph{loop=true}

$\qquad$\textbf{$\qquad$while}\emph{ loop}

$\qquad\qquad\qquad$\textbf{set $\ell=\ell+1$}

$\qquad\qquad\qquad$\textbf{set} $\beta_{k,n}=\eta_{\ell}$

$\qquad\qquad\qquad$\textbf{set} $z=y^{k,n}+\beta_{k,n}v^{k,n}$

$\qquad\qquad\qquad$\textbf{if }$\phi\left(z\right)$\textbf{$\leq$}
$\phi\left(y^{k}\right)$\textbf{ then}

$\qquad\qquad\qquad\qquad$\textbf{set }$n$\textbf{$=$}$n+1$

$\qquad\qquad\qquad\qquad$\textbf{set }$y^{k,n}$\textbf{$=$}$z$

$\qquad\qquad\qquad\qquad$\textbf{set }\emph{loop = false}

$\qquad$\textbf{set }$y^{k+1}$\textbf{$=$}$\boldsymbol{A}_{C}\left(y^{k,N}\right)$

$\qquad$\textbf{set }$k=k+1$ \medskip{}
 \label{alg1}
\end{algorithm}

In general, the Superiorized Version of the Basic Algorithm outputs
solutions that are essentially as constraints-compatible as those
produced by the original (not superiorized) Basic Algorithm. However,
due to the repeated steering of the process toward reducing the value
of the target function $\phi$, we can expect that the output of the
Superiorized Version will be superior (from the point of view of $\phi$)
to the output of the original algorithm.

In recent years, the algorithmic structure of the superiorization
method has undergone some evolution in ways that offer benefits in
pCT. The details of this evolution can be found in the Appendix of
\cite{YC17}, titled ``The algorithmic evolution of superiorization''.

A comprehensive overview of the state of the art and current research
on superiorization appears in our continuously updated bibliography
Internet page that currently contains 138 items \cite{sup-bib}.
Research works in this bibliography include a variety of reports ranging
from new applications to new mathematical results on the foundations
of superiorization. A special issue entitled: ``Superiorization:
Theory and Applications'' of the journal Inverse Problems \cite{Sup-Special-Issue-2017}
contains several interesting papers on the theory and practice of
SM. A special issue entitled: ``Superiorization versus Constrained
Optimization: Analysis and Applications'' of the Journal of Applied
and Numerical Optimization (JANO) appeared in 2020 \cite{Sup-Special-JANO}.

A word About the history: The superiorization method
was born when the terms and notions \textquotedblleft superiorization\textquotedblright{}
and \textquotedblleft perturbation resilience\textquotedblright ,
in the present context, first appeared in the 2009 paper of Davidi,
Herman and Censor \cite{DHC09} which followed its 2007 forerunner
by Butnariu et al. \cite{BDHK07}. The ideas have some of their roots
in the 2006 and 2008 papers of Butnariu et al. \cite{BRZ2006} and
\cite{BRZ2008}. All these culminated in Ran Davidi\textquoteright s
2010 PhD dissertation \cite{davidi2010} and the many papers since
then cited in \cite{sup-bib}.

\subsection{Derivative-free Superiorization (DFS) with Component-wise Perturbations}

Superiorization reduces, but not necessarily minimizes, the value
of a target function while seeking constraints-compatibility. When
the perturbation steps are computationally efficient, the superior
result is obtained with essentially the same computational cost as
that of the original feasibility-seeking algorithm.

In the literature on superiorization, the perturbations that interlace
target function reduction steps into the basic algorithm have, up
to now, been done mostly by using negative gradients (or subgradients)
directions and, thus, require some form of differentiability of the
target function. In \cite{CHS2019}, we introduced \textit{component-wise}
perturbations that allow local non-ascent of the target function.
These enable the SM to be applied with target functions that are \textit{not}
differentiable, similarly to such situations in optimization theory,
where coordinate descent approaches are used.

The ramification of DFS for practical applications of the SM to IMPT
treatment planning are meaningful. For example, normal tissue complication
probability (NTCP), which is a predictor of radiobiological effects
for organs at risk, should be used as an objective function in the
mathematical problem modeling and the planning algorithm. Using NTCP
or similar functions is hampered because these functions are, in general,
empirical functions whose derivatives cannot be calculated, see, e.g.,
\cite{Niemierko2007}. In a recent paper \cite{Nystrom2020} the authors
say that ``...practical tools to handle the variable biological efficiency
in Proton Therapy are urgently demanded...'' highlighting this need
for the near future, but also in a longer perspective, of the proton
therapy community.

By considering component-wise perturbations, we generalized previous
superiorization schemes to enable the use of a more extensive selection
of methods for step-wise reduction of the target function. As a first
step in validating component-wise perturbations, we presented in \cite{CHS2019}
a new superiorization scheme for reducing total variation (TV) during
image reconstruction from projections. More recently, this work has
been continued in \cite{CensorDerivativeFree2019arXiv}.
In that paper, DFS is put in context with the large field of derivative-free
optimization (DFO) with many relevant references and a tool, called
a ``proximity-target curve'', for deciding which of two iterative
methods is \textquotedblleft better\textquotedblright{} for solving
a particular problem is developed. It is worthwhile to mention that
other methodologies such as averaged stochastic gradient descent and
automatic differentiation, that are widely used in optimization theory,
in particular in the field of machine learning, also cope with lack
of differentiability and employ ``coordinate-wise'' searches. There
is certainly a potential to learn from the advances in these fields
toward improving DFS.

\subsection{The Fully-Discretized Problem Formulation of pCT Image Reconstruction}

\label{sec:GenpCT} The water equivalent path length (WEPL) of a proton
through an object, i.e., the length of the path the proton travels
through water that leads to the same mean energy loss as in the object,
can be expressed as a line integral of relative (to water) stopping
power (RSP) along the path \cite{HSBWG12}. In practice, the object
is described by a set of basis functions, and the integrals are expressed
as discrete sums. This leads to a linear system $Ax=b,$ where the
$ij$th entry of the matrix $A$ is the intersection length of the
path of the $i$th proton with the $j$th voxel (basis function) and
the $i$th component of the measurement vector $b$ is the $i$th
proton's measured WEPL. The mathematical formulation of the fully-discretized
pCT reconstruction problem is: Given $\ensuremath{A}$ and $b$, estimate
$\ensuremath{x}$.

The system matrix $A$ is generated by calculating the path length
through each basis function (normally a voxel). Different from the
straight-line assumption in x-ray CT, in pCT, discrete steps along
most likely path (MLP) approximations developed in \cite{SPTS08},
created by multiple small-angle Coulomb-scattering of protons in the
object, are used. Thus, proton paths through the reconstructed object
are approximated at a finite set of points along each MLP. Employing
a linear path approximation through individual basis functions simplifies
the calculation. It is justified by the minor deviation of a proton
path from a straight line on the scale of the voxel dimension ($1$-$2$
mm). However, the need to identify intersection lengths of protons
through individual voxels creates high computational demand with the
need to identify intersection lengths of millions of protons through
millions of voxels. We addressed this problem by using GPGPUs (general-purpose
graphics processing units) that perform calculations that would otherwise
typically be conducted by the CPU (central processing unit). Such
GPGPUs were used for the implementations of pCT reconstruction codes
(see Subsection \ref{sec:GPGPU} below).

\subsection{A New Algorithm for TV-Superiorized (TVS) pCT Image Reconstruction}

\label{sec:NTVS}The efficacy of the SM for image reconstruction in
pCT has been shown in previous work \cite{PSCR10}. In that work,
we superiorized the total variation (TV) as the target function to
improve pCT image quality. The usefulness of TVS was demonstrated
for pCT image reconstruction with different superiorized versions
of the block-iterative diagonally relaxed orthogonal projections (DROP)
algorithm \cite{DROP-2008}. Two TVS schemes added-on to DROP were
investigated; the first carried out the superiorization steps once
per cycle and the second once per block.

A new version of the TVS algorithm, referred to as NTVS (New TVS),
was investigated in \cite{BS2019}. Compared to the original TVS algorithm
published by Penfold et al. in \cite{PSCR10}, the NTVS includes several
structural changes and new aspects previously not investigated. It
combines properties that were scattered among previous works on TVS
in x-ray CT but were never combined in a single algorithm, neither
for x-ray CT nor for pCT. These properties are: (1) exclusion of the
TV reduction verification step; (2) usage of powers of a perturbation
kernel $\alpha$ to control the step-sizes $\beta_{k}$ in the TV
perturbation steps; (3) incorporation of the user-chosen integer $N$
that specifies the number of TV perturbation steps between consecutive
feasibility-seeking iterations; and (4) incorporation of a new formula
for calculating the power of $\alpha$ that is used to calculate the
step-size of the perturbation steps. The notation and other algorithmic
details of the NTVS algorithm can be found in \cite{BS2019}.

\subsection{Adding Robustness to pCT Image Reconstruction}

Imaging systems commonly contain some uncertainty in modeling and
measurement. For pCT, there is uncertainty in the estimated MLP and
the measurement of the WEPL of individual protons. In standard algorithms,
this uncertainty, often measured by the residuals $r=\parallel b-Ax\parallel$,
is considered part of the measurements $b$ rather than the modeled
system $Ax$. In robust systems, errors are modeled as part of both
the system matrix $A$ (in pCT, intersection length of individual
proton MLPs with individual voxels) and the measurement vector $b$
(in pCT, the WEPL). Methods such as total least squares,
ridge regression, and Tikhonov regularization are classic techniques
for robust systems, and all can be solved as a diagonal perturbation
to the normal equations of least squares, see for example \cite{GoluVanl96}.
The main difference relates to the regularization parameter, which
is typically a small number related to the uncertainty parameter,
and bounds on $A$ and $b$. A more recent technique, Bounded Data
Uncertainty (BDU) models the uncertainty as perturbations $E$, to
the matrix $A$, such that the actual system is $(A+E)x$. The perturbations
are bounded such that $\|E\|\leq\eta$. For the worst case perturbations
\cite{minmax} and \cite{watson} and best case perturbations \cite{minmin}
and \cite{degenminmin} the resulting problem is solved by $x=(A^{t}\Phi A+\Psi)^{-1}A^{T}\Phi b$,
which again can be solved by a diagonal perturbation to the normal
equation of least squares. In BDU, the perturbation to the diagonal
is dependent on the solution and must be solved for iteratively using
a ``secular equation,'' which is a one dimensional, nonlinear equation
involving uncertainty parameter $\eta$, and bounds on $A$ and $b$.
These diagonal perturbations can be thought of as perturbing the singular
values of the underlying system of $A$ and $b$. Analogously, in
\cite{doi:10.1063/1.5127700}, we introduced the \emph{fully-simultaneous
adaptive iterative solver} (FSAIS) for the fully-discretized
problem formulation of pCT image reconstruction and provided a convergence
analysis. To obtain a sparse robust solution, we formed the augmented
linear system

\begin{equation}
\left[\begin{array}{cc}
-\varPsi & A^{T}\\
A & \varPhi^{-1}
\end{array}\right]\left[\begin{array}{c}
x\\
r
\end{array}\right]=\left[\begin{array}{c}
0\\
b
\end{array}\right],\label{eq:FAISa}
\end{equation}
where $A$, $x$, and $b$ are defined as in \ref{sec:GenpCT}, and
the choice of the matrix parameters $\varPsi$ and $\varPhi$ is discussed
below. Solving the top equation for $x$ and the second equation for
$r$, one obtains the following system of equations:

\begin{equation}
r=\varPhi\left(b-Ax\right),\label{eq:FAISb}
\end{equation}

\begin{equation}
x=\varPsi^{-1}A^{T}r,\label{eq:FAISc}
\end{equation}
which reduces to
\begin{equation}
x=\varPsi^{-1}A^{T}\varPhi\left(b-Ax\right).\label{eq:FAISd}
\end{equation}

This equation opens a formal way of selecting an iterative algorithm
that leads to robust RSP values. Given a current iteration vector
$x^{k}$, the FSAIS generates the next iteration vector $x^{k+1}$
by

\begin{equation}
x^{k+1}=x^{k}+\varPsi^{-1}A^{T}\varPhi(b-Ax^{k}).\label{eq:FSAIS}
\end{equation}
Since many robust methods result in diagonal perturbations
with values that typically lie in a similar area, in FSAIS the two
diagonal parameter matrices $\varPsi$ and $\varPhi$ are free parameters
that can be adjusted for robustness and speed of convergence. The
parameter $\Phi$ handles normalization and weighting between the
rows of $A$ and $b$. The parameter $\Psi$ includes the relaxation
paramter and adjusts weighting between the uncertainty $\eta$, the
elements of $x$, and the matrix $A$. Our choices for these parameters
are $\varPhi_{i}:=\frac{{1}}{\|A_{i}\|^{2}}$, where $A_{i}$ is the
$i$th row of the matrix $A$, and $\varPsi_{j}^{-1}(k):=(1-x_{j}^{k})\lambda(k)$,
where $\lambda(k)$ is the relaxation parameter at the $k$th iteration
and $x_{j}^{k}$ is the $j$th component of the $k$th iteration vector.
In a pCT reconstruction study with the Catphan\textsuperscript{®}
CTP404 sensitometry module (The Phantom Laboratory, Inc., Salem, NY),
which contains cylindrical inserts with a wide range of RSP, FSAIS
was shown to obtain results similar to the DROP algorithm of \cite{DROP-2008}
for RSP values near $1$ but obtained significantly better results
than DROP for denser materials (RSP\textgreater 1) such as acrylic
(1.160), Delrin\textsuperscript{®} (1.359), and Teflon\textsuperscript{®}
(1.790). FSAIS was also able to handle missing data, i.e., proton
histories in an angle interval being removed. This was done to simulate
situations, for example, where protons cannot penetrate individual
body sections because the proton energy would not be sufficient. Using
the filtered back projection (FBP) reconstruction along straight proton
paths as the initial iterate, FSAIS stayed within 2\% of the RSP for
up to 60 degrees of lost data and was within 1\% for almost all materials
for 30 degrees of lost data.

\subsection{Development of a Motion-Adapted Iterative Reconstruction Algorithmic
Framework for pCT}

In the past and up to now, it was assumed that the object that undergoes
pCT imaging is static throughout the pCT data acquisition. However,
this is not the case when imaging, for example, the lungs because,
during the image acquisition, the acquired data belong to different
motion states at different specific time instances. It is important
to note here that we register individual protons with a time tag during
the data acquisition process. During the data acquisition, one can
also measure a surrogate breathing signal that gives information about
the tidal volume of the lungs. Common concepts in four-dimensional-CT
(4D-CT) imaging with the registration of the breathing motion are
those of a ``reference state'' and the ``reference image'' which
represent the patient lungs in a specific state of filling and, typically,
at rest, respectively, for example, at the end of normal expiration
when the breathing airflow ceases for a moment. In
the work-in-progress described next, we developed and investigated
an algorithmic framework that enables the reconstruction of the reference
image of a breathing patient assuming that the tissue motion can be
described by accurate deformation vector fields, see, e.g., \cite{Thomas2014}.
For ordinary (motion-free) pCT data acquisition, one uses a detector-fixed
coordinate system for tracking discrete proton coordinates for MLP
calculation and an object-fixed coordinate system for reconstructing
the RSP; both coordinate systems are represented by a regularly spaced,
three-dimensional voxel grid \cite{Schultze2021}. When motion is
present, the object-fixed voxel grid is only regular for the reference
state. However, it is deformed for all other motion states according
to the deformation described by the motion model. The general reconstruction
equation $Ax=b$ is retained because motion only influences the $A$-matrix
elements according to the MLP intersection length of the deformed
grid cells. Assuming that the RSP values of the deformed grid cells
are retained (an approximation), the reconstructed image corresponds
to the reference image when the solution vector is assigned to the
rectangular voxel grid, and any other motion state when it is assigned
to the deformed grid of that motion state.

\begin{figure}[t]
\centerline{\includegraphics[width=0.95\linewidth]{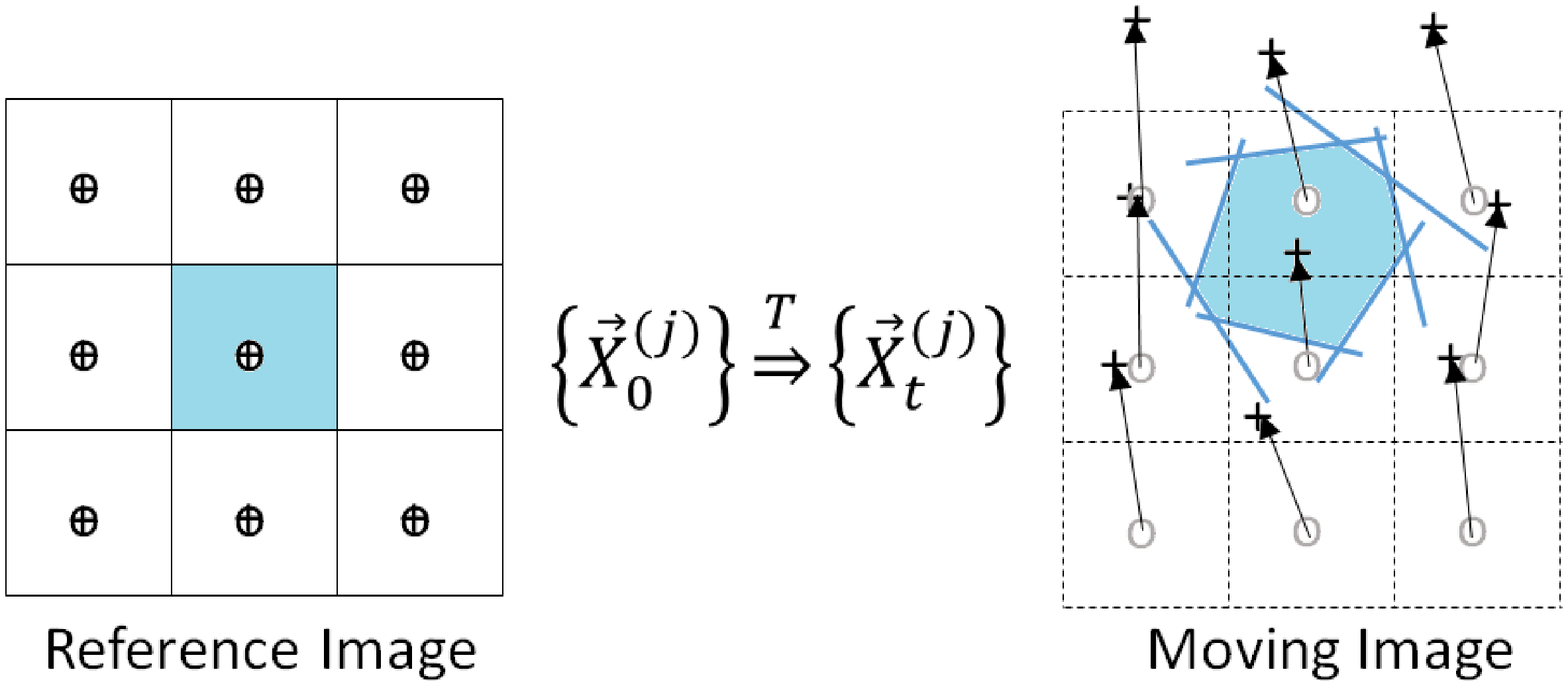}}
\caption{Two-dimensional schematic of the reference image (left) with a regular
system of points marking the centers of an object-centered fixed voxel
grid $j=1,2,\ldots,J$. In the moving image (right), the position
of the original voxel center points have shifted as a whole (translation)
and relative to each other (deformation) to new locations due to tissue
motion, which is described by a deformation vector field $T$. The
blue-shaded regions are Voronoi cells (see text for details).}
\label{figDVF}
\end{figure}

We define a ``motion model,'' in the context of pCT image reconstruction,
to be a system (like a blackbox) capable of describing the three-dimensional
(3D) spatial displacement of human moving tissues at any time instance
relative to their reference position. To make this statement more
concrete, we consider a 3D regular voxel grid fixed in the object-fixed
coordinate system whose voxel centers are, for each voxel $j$, at
the 3D position vector $\overrightarrow{X}_{0}^{(j)}.$ For every
voxel index, $j=1,2,\ldots,J,$ the pair $(y_{0}^{(j)},\overrightarrow{X}_{0}^{(j)})$
represents the relative stopping power (RSP) of the voxel $j$, denoted
by $y_{0}^{(j)},$ and the location $\overrightarrow{X}_{0}^{(j)}$
of the voxel center in the reference image, see Figure \ref{figDVF}.

According to the modeled displacement of the object tissues due to
motion, at time $t$, the pair $(y_{0}^{(j)},\overrightarrow{X}_{0}^{(j)})$
becomes $(y_{0}^{(j)},\overrightarrow{X}_{t}^{(j)}),$ meaning that
the RSP value $y_{0}^{(j)}$ remains unchanged, but its reference
point has moved to a new spatial location $\overrightarrow{X}_{t}^{(j)},$
not necessarily conforming with the original fixed grid, as determined
by the nature of a transformation $T$ described by a deformation
vector field that is, in general, not repeatable nor periodic and
is described by the motion model.

Our motion-adapted pCT reconstruction makes use of the concept of
Voronoi diagrams and cells, see, e.g., \cite[Subsection 5.5]{preparata2012computational},
\cite{Berg2008}. We use the following definition.

\begin{definition} \label{def:Voronoi_diagram} 
Let $P:={p_{1},p_{2},\ldots,p_{J}}$ be a set of $J$ voxel center
points (called sites) in the 3D reference image space and let $\text{dist}(p_{j},p_{k})$
denote the Euclidean distance between voxel center point $p_{j}$
and any other voxel center point $p_{k},k\neq j$. The ``Voronoi
diagram'' of $P$, is defined as the partition of the 3D space into
$J$ connected 3D cells (called ``Voronoi cells'' or ``Voronoi
polygons'') such that any point $q$ in the image space belongs to
the Voronoi cell corresponding to a site $p_{j}$ if and only if $\text{dist}(q,p_{j})<\text{dist}(q,p_{k})$
for each $p_{k}\in P$ with $k\neq j$. \end{definition}

It is understandable from Figure \ref{figDVF}, when expanded into
the third spatial dimension, that boundaries of Voronoi cells are
formed by plane sections that are created by the intersection of bisecting
half-spaces between pairs of sites, line segments formed by the intersection
of these planes, and vertices formed by the intersection of two or
more of these line segments. The ``bisecting plane'' of two points
$p$ and $q$ is the plane that is perpendicular to the line segment
$\overline{pq}$ and intersects it at the midpoint. This bisecting
plane divides the 3D space into two half-spaces. Then the Voronoi
cell of point $p_{k}$, denoted by $\mathcal{V}(p_{k})$, is

\begin{equation}
\mathcal{V}(p_{k})=\bigcap_{j\neq k,j=1}^{J}h(p_{k},p_{j}),\label{eq:Voronoi_cell}
\end{equation}

\label{def:Voronoi_cell}

\noindent where $h(p_{k},p_{j})$ is\textbf{ }the bisecting plane
of the points $(p_{k},p_{j})$.

\begin{figure}[t]
\centerline{\includegraphics[width=0.75\linewidth]{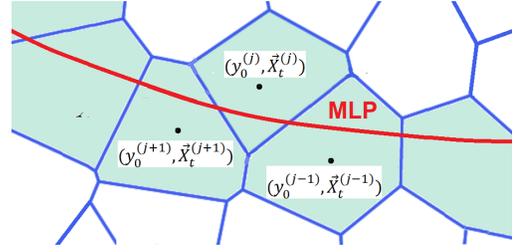}}
\caption{Two-dimensional schematic of the pCT reconstruction problem of a moving
organ at a specific time instance $t$ when the organ is represented
by the Voronoi cell basis functions shown. The most likely path (MLP)
of an individual proton intersects the shaded cells. The summation
of the intersection lengths of the $i$th MLP with the $j$th voxel,
$a_{ij}$ multiplied by the unknown RSP values $y_{0}^{j}$ of each
cell is set equal to the $i$th component of the $b$ vector of measured
WELP values, and thus form a linear system of equations $Ay_{0}=b$
that is solved iteratively for the reference image vector $y_{0}$
by a feasibility-seeking algorithm.}
\label{figVD1}
\end{figure}

In our approach to motion-adapted pCT reconstruction, which is illustrated
in Figure \ref{figVD1}, the Voronoi cells become the new basis functions,
and the originally formulated fully-discretized problem of pCT reconstruction
remains unchanged. The specific Voronoi cell basis functions depend
on the motion state of the tissues, which is described by a motion
model. The motion model allows calculation of the displacement vector
for individual voxel points at any given instance of time. The new
positions of the voxel points are used to compute the plane segments
of the Voronoi cells. Together with an MLP algorithm, one can then
calculate the intersection length of the $i$th proton through the
$j$th Voronoi cell basis function, and the $i$th component of the
WEPL measurement vector $b$ gives the $i$th proton's measured WEPL.
This results in a linear system of equations that is solved iteratively
by a feasibility-seeking algorithm. The unknown RSP values of each
cell are assumed not to change from their values in the reference
image. Thus, the reconstructed voxel image where each voxel is assigned
the reconstructed RSP value $y_{0}^{(j)},j=1,2,\ldots,J,$ corresponds
to the reference image. One can also calculate the image at any point
in time by calculating a weighted average of the RSPs with weights
that are proportional to the area a voxel shares with overlapping
Voronoi cells.

The motion model parameters could be known at the time of pCT imaging
from a separate motion imaging study, or they could be included in
the image reconstruction problem and estimated after the imaging.
By way of example only, we mention here the lung motion model originally
developed by Low and colleagues \cite{low05,low13}. This model has
recently been applied in conjunction with an iterative projection
algorithm for a motion-compensated simultaneous algebraic reconstruction
technique (MC-SART) of cone beam CT (CBCT) \cite{Guo2019,Chee2019}.
There is, however, a fundamental conceptual difference between our
approach and the MC-SART algorithm. To ``handle'' the motion, the
authors modified the reconstruction algorithm (SART in their work)
while we, to handle the motion, modify the problem (the linear system
of equations in our work). Our approach is, therefore, general in
the sense that once the problem is appropriately modified to handle
the motion, one can apply any iterative algorithm to the ``modified
problem.'' Our framework can take the output of any motion model
as input to the algorithm for computation of the basis functions at
each instance in time when proton histories are acquired. The motion
model and the algorithm are, thus, two separate entities that have
to be used in tandem when applied to simulated or real data.

Besides the motion model of Low et al., other motion models exist,
see, e.g., \cite{Eom2010,Ehrhardt2011,Zhang2013}.

\subsection{Blob Basis Functions for pCT Image Reconstruction}

``Blob'' basis functions, called ``radial basis
functions'' in approximation theory, see, e.g., \cite{blobs2003},
are spherically symmetric basis functions for CT reconstruction that
were suggested in the field of image reconstruction from projections
by Lewitt in the early 1990s \cite{Lewitt_1992,Lewitt_1990}. These
blobs are generalizations of a well-known class of functions used
in digital signal processing called Kaiser-Bessel window functions
and they yield excellent results in multiple imaging modalities, see,
e.g., the enlightening tutorial by Herman \cite{herman_2015}. Matej
and Lewitt \cite{Matej1995,Matej1996} provided a careful investigation
of how the blob basis functions should be chosen when they are used
in the context of image reconstruction from projections. Since then
blobs have been used extensively for image reconstruction in X-ray
computerized tomography, positron emission tomography, single photon
emission computerized tomography, optoacoustic tomography and electron
microscopy (consult \cite{herman_2015} for references and details).
Adhering to the term ``blobs'', as commonly used in this field,
we have started to develop a methodology for using blobs instead of
standard square voxels in pCT. The use of blobs will potentially improve
pCT image quality and computational efficiency.

Current pCT makes use of conventional voxels for reconstructing and
representing reconstructed images. However, proton paths through the
reconstruction object are curved and must be approximated at a finite
set of points along each path. The computational burden to identify
intersection lengths of protons through individual voxels is quite
large, especially when the object elements are not spatially invariant.
Using blobs as a replacement of voxels in pCT can significantly reduce
this computational burden because of the spherical symmetry of the
blobs. We presented initial encouraging results in collaboration with
UCLA graduate student Howard Heaton as a poster at the 2016 Joint
Mathematics Meetings (JMM) \cite{poster-jmm}.

\begin{figure}[t]
\centerline{\includegraphics[width=0.9\linewidth]{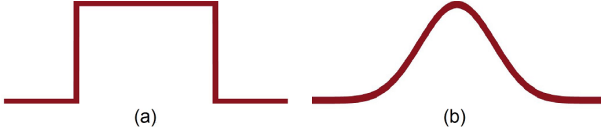}}
\caption{Profile of a voxel (a) and a blob (b).}
\label{figblobs1}
\end{figure}

Voxels have a uniform value inside a set domain, and blobs are spherically
symmetric and taper smoothly to zero at their border, see Figure~\ref{figblobs1}.
The formal definition for blob basis functions for CT imaging, first
proposed by Lewitt \cite{Lewitt_1990}, is as follows:
\begin{equation}
b_{j}(r):=\begin{cases}
\frac{\big[1-(r/a)^{2}\big]^{m/2}}{I_{m}(\alpha)}I_{m}\big(\alpha\sqrt{1-(r/a)^{2}}\big), & \text{if \ensuremath{r\in[0,a],}}\\
0, & \text{otherwise},
\end{cases}
\end{equation}

\noindent where $b_{j}$ is the basis function of the $j$th blob,
$r$ is the radial distance from the center of the $j$th blob, $I_{m}$
denotes the modified Bessel function of the first kind of order $m$,
$a$ denotes the radius of the blobs, and $\alpha$ is a nonnegative
real number that controls the shape and taper of the blob. It is common
practice to choose $m=2$, which gives a smooth and differentiable
blob function. Following Benkarroum et al. \cite{Benkarroum_2015},
we have chosen the blob parameters to be $a=2.453144$ mm and $\alpha=13.738507$.
Those parameter values were chosen for optimal representation of piecewise-constant
images.

To generate the system matrix $A$, the path length through each basis
function must be computed. Employing a linear path approximation through
individual basis functions (voxels or blobs) allows the use of the
Radon transform $\mathcal{R}$ to approximate each entry ${A}_{ij}$,
i.e.,
\begin{equation}
{A}_{ij}=\int_{path}b_{j}(\vec{s}\hspace*{5pt})\ d\vec{s}\ \approx\ [{\mathcal{R}}b_{j}](\ell_{i},\theta_{i}),
\end{equation}

\noindent where
\begin{equation}
[\mathcal{R}b_{j}](\ell_{i},\theta_{i})=\int_{line}b_{j}\left(\sqrt{\ell_{i}^{2}+z^{2}},\ \theta_{i}+\tan^{-1}\left({z}/{\ell_{i}}\right)\right)\ dz
\end{equation}
is the Radon transform of the straight-line approximation of the path
through the blob. Note that 
with blobs, the Radon transform depends solely on perpendicular distance
from the blob center, $\ell_{i}$.


To construct the system matrix, we use the following problem formulation.
\begin{problem}
\label{prob:blob-algo}\textbf{Generation of the $A$-matrix for blob
basis functions}: Let $\{\vec{\varphi}^{\hspace{2.5pt}\ell}\}_{\ell=1}^{L}$
denote an ordered set of points along a proton path in $\mathbb{R}^{3}$.
Suppose that this path passes through an object represented with a
set $\{b_{j}\}$ of blob basis functions. Using successive points
$\vec{\varphi}^{\hspace{2.5pt}\ell}$ along the path, uniquely estimate
each nonzero blob intersection length of the path to generate the
corresponding system matrix.
\end{problem}
A key step in the algorithm presented in \cite{poster-jmm} is to
identify blobs within a given proximity of a point $\vec{\varphi}^{\hspace{2.5pt}\ell}$
on the path. This is accomplished by identifying a corresponding point
$\vec{\varPhi}^{\hspace{2.5pt}\ell}$ located nearby on the grid of
blob centers with grid unit $\beta.$ 
The algorithm identifies each nonzero blob intersection length by
identifying a blob in proximity to $\vec{\varphi}^{\hspace{2.5pt}\ell}$.
Then the algorithm begins to cycle through nearby blobs within a chosen
range $\kappa$. Finally, it restricts intersection lengths to being
assigned during the last step the proton takes before passing the
blob center in depth. Further details can be found in \cite{poster-jmm}.

\subsection{Algorithms for IMPT: Dose-volume Constrained Split Feasibility}

IMPT is a rapidly developing field wherein active pencil beam scanning
became the norm in proton and ion beam treatment centers. IMPT requires
scanning of a particle treatment field with a grid of pencil beam
spots that are rapidly and often repeatedly ``visited'' during the
treatment to deliver a spot dose with varying intensity. Such a scan
demands fast IMPT algorithms that solve the inverse problem with multiple
dose constraints within target volumes and organs at risk. During
our research, we tackled the additional complication that dose constraints
are often modified, by the treatment planner, allowing violation of
the constraints by a specific percentage of the dose for a specific
percentage of the volume. The method we developed to handle this is
also useful for x-ray IMRT, see \cite{Penfold2017}.

The linear feasibility problem (LFP) formulation, which is a special
case of the well-known convex feasibility problem (CFP), see, e.g.,
\cite{bb96}, forms a basic model for the inverse problem in the fully-discretized
approach to both IMPT and IMRT treatment planning. In 2015, we showed
that IMPT inverse planning is possible by using a fully-discretized
model and a feasibility-seeking algorithmic approach \cite{Penfold2015}.
In particular, we demonstrated on a simple 2D example that solutions
meeting the planning objectives could be found by these feasibility-seeking
iterative projection algorithms.

In the planning of IMPT or IMRT, one uses dose-volume constraints
(DVCs) to evaluate treatment plans. Mathematically, DVCs are percentage-violation
constraints (PVCs). PVCs single out individual subsets of the existing
volume constraints and allow a specified percentage of dose constraints
to be violated in each subset. Without incorporating the PVCs into
the mathematical inverse planning model and algorithm themselves,
it is not possible to guarantee that an appropriate solution will
be found \cite{Penfold2017}.

A tractable model and an algorithmic approach to solve the IMPT/IMRT
inverse planning problem as a feasibility problem that includes DVCs
has been developed. A rigorously defined notion of PVCs cause integers
to enter the problem, which makes it difficult to solve. To circumvent
this difficulty, we reformulated the PVC with the aid of a ``sparsity-norm''
that counts the number of nonzero entries in a vector.
There is a rich literature on how to handle this norm, called also
the $\ell_{0}$-norm, which is still stimulating recent and ongoing
related work. It is not in the scope of this paper to review this
field but see, e.g., \cite{hesse-luke-2014} and its references, where
the problem is formulated and handled as a full fledged constrained
optimization problem. Using the $\ell_{0}$-norm enables to enforce
the DVCs and leads to redefining the ``linear feasibility problem
with PVCs'' as another feasibility problem that includes non-convex
constraints for the sparsity-norm.
\begin{problem}
\label{prob:imrt+dvc}\cite{Penfold2017}\textbf{ Linear Interval
Feasibility with DVC for the inverse problem in the fully-discretized
approach to IMRT treatment planning}. Find $x^{\ast}\in R^{n}$ for
which
\begin{gather}
0\leq A_{1}x\leq(1+\beta)b^{1},\label{eq:dvc1}\\
b^{3}\geq A_{2}x\geq b^{2},\label{eq:dvc2}\\
0\leq A_{3}x\leq b^{4},\label{eq:dvc3}\\
x\geq0,\label{eq:dvc4}\\
\Vert(A_{1}x-b^{1})_{+}\Vert_{0}\leq\alpha m_{1},\label{eq:dvc5}
\end{gather}
where $A_{1}\in R_{+}^{m_{1}\times n}$, $A_{2}\in R_{+}^{m_{2}\times n}$,
$A_{3}\in R_{+}^{m_{3}\times n}$ are given matrices, $b^{1}\in R_{+}^{m_{1}}$,
$b^{2},b^{3}\in R_{+}^{m_{2}}$, $b^{3}\in R_{+}^{m_{3}}$ are given
vectors, and $\beta>0$ and $\alpha\in\lbrack0,1]$ are given real
numbers.
\end{problem}
In this problem, applicable to both IMPT and IMRT, the sparsity constraint
(\ref{eq:dvc5}) takes place in the space $R^{m_{1}},$ where the
vectors of doses in the organs at risk reside. Therefore, we cannot
use plain feasibility-seeking methods but need feasibility-seeking
methods for ``split feasibility problems.'' Thus, we recognized
that Problem \ref{prob:imrt+dvc} is a split feasibility problem.
Split feasibility problems were introduced first in \cite{ce94} and
further studied in \cite{mssfp,cgr} and many other publications.
The reader may consult the brief review of ``split problems'' formulations
and solution methods in \cite{BCG2019} for more details and references.
For the solution of Problem \ref{prob:imrt+dvc}, which includes a
non-convex sparsity-norm induced constraint, we developed a new iterative
projection algorithm, which is a combination of the $CQ$-algorithm
\cite{Byrne2002} and the automatic relaxation method (ARM) \cite{arm}.
Full details of this approach appear in \cite{Penfold2017}, which
applied it to a single OAR. Following this line of development, we
expanded the model to include DVCs for several OARs in a more recent
paper \cite{BCG2019}. Handling DVCs in radiation
therapy treatment planning is a viable research area and a variety
of techniques have been applied to solve it, see, e.g., \cite{MKA2020}
and references therein.

\section{Development of software tools and phantoms to test advanced pCT and
IMPT algorithms}

\label{sec:tools-phantoms}The demand for high-performance computing
capabilities for pCT image reconstruction and IMPT algorithm development,
is apparent. In this section, we summarize developments facilitating
fast algorithms for pCT and IMPT.

\subsection{GPGPU Data Partitioning}

\label{sec:GPGPU}An essential aspect of our work as members of the
pCT collaboration has been to provide tools for fast and efficient
image reconstruction using GPGPU computations. The pCT detector hardware
was developed in two phases from a fairly small and slow detector
system in Phase I to a much faster Phase II system completed in 2015
\cite{BJSS16}. As the reconstructed image gets larger, the number
of proton histories that enter the reconstruction is up to one hundred
times the number of voxels in the image, see \cite{WSS12}. Currently,
objects requiring a six minutes scan time (up to 360 million proton
histories) with the ``Phase II proton CT scanner'' can be reconstructed
on the order of a few minutes \cite{SKGPSS15,JBCGK16,8532871}. As
we image larger objects with protons, not only does the reconstruction
time grow significantly, but, more importantly, the memory requirements
grow too. Therefore, we need to split the data into coherent data
blocks for efficient reconstruction. Since the Phase II pCT scanner
acquires the data during a 360-degree rotation of the object relative
to the pCT scanner with a wobbled or scanned beam spot \cite{JBCGK16,JBDGH16},
we can efficiently split the data into slices orthogonal to the rotational
axis, see \cite{8532871}.

For a nominal proton entrance energy of 200 MeV and a few tens of
MeV exit energy, proton path histories within three standard deviations
of the nominal path in lateral direction do not pass through more
than three 1 mm slices. So, when reconstructing a slice, no more than
three slices above or below the slice need to be available in memory
of a particular GPGPU to reconstruct the image efficiently, see \cite{PK}.
When reconstructing slice-by-slice, we can allocate up to one GPGPU
per slice. For proton histories passing through multiple slices, each
GPU associated with a slice holds the most recent copy of the RSP
values of adjacent slices. Updates from adjacent slices are transferred
when the thread calculating the update finishes an iterative cycle
of the algorithm.

The results presented in \cite{8532871,PK} show that this method
does not cause a significant image quality reduction while permitting
reconstruction of the anthropomorphic head phantom described below
in under 10 minutes on NVIDIA P100 GPGPUs. Further speedup to reconstruction
in under 1 minutes is within reach.

\subsection{Monte Carlo Simulation Tools}

\label{sec:MC-simu-tools} Monte Carlo (MC) simulations are an essential
tool in the development and testing of algorithms for pCT image reconstruction
and serve as a forward calculation tool for the radiation dose deposited
in object voxels of IMPT plans. While testing algorithms on real-world
experimental data is equally important, MC-based data are more readily
available and provide a great deal of flexibility when testing and
developing new algorithms or algorithmic variants.

We have made extensive use of the general-purpose MC simulation tool
Geant4 \cite{geant4} to develop a faithful model of the Phase II
pCT detector system developed by the pCT collaboration (see Section
\ref{sec:GPGPU}). For this project, we reproduced the exact geometry
and composition of materials of the scanner as well as the research
proton beamline at the Loma Linda University Medical Center. Details
of the initial Geant4 platform development and the methods we used
to validate its output, in connection with reconstruction algorithms
that we developed, can be found in Dr. Giacometti's thesis publications,
see \cite{GiacomettiDis}. We then used pCT simulations produced by
that platform in addition to experimental data to test the performance
of the new TVS algorithms described in Section \ref{sec:SM}.

Later, we implemented the Phase II scanner in the TOPAS (\textbf{TO}ol
for \textbf{PA}rticle \textbf{S}imulation) platform, which is also
based on the Geant4 toolkit but has a more user-friendly interface
\cite{PSSFP2012}. We also tested the Phase II pCT scanner at the
Northwestern Medicine Chicago Proton Center (NMCPC), where it is operated
on a clinical proton beamline. The TOPAS model of the NMCPC beamline
was validated with beam profile measurements acquired at NMCPC and
by acquiring and simulating pCT images using a wobbled 200 MeV proton
beam. With the TOPAS tool, we then investigated the sources of systematic
uncertainty introduced during the steps of iterative RSP reconstruction
with the DROP-TVS algorithm \cite{PSCR10} for CT QA phantom modules
(Catphan\textsuperscript{®}, The Phantom Laboratory Incorporated,
Salem, NY, USA) and a digital representation of a pediatric patient
with known ground truth RSP values in the simulation, see \cite{PRMGBSF2017}.
The important point here is that with Monte Carlo simulations of the
pCT system performance, we can intentionally include or exclude different
sources of random and systematic uncertainties that arise from random
fluctuations in the proton energy loss and scattering, which are unavoidable,
and sources of imperfections in the beam source or the detector construction.
For example, we had to understand the origin of ring artifacts seen
in the reconstruction of homogeneous phantoms. Through systematic
Monte Carlo studies, we learned that these artifacts were related
to using a WEPL calibration phantom with discrete thickness steps.
A simulation of the pCT detector replacing the step phantom with a
wedge phantom demonstrated that the ring artifacts could be significantly
reduced \cite{PRMGBSF2017}.

At the end of 2016, we shipped the Phase II pCT scanner to the Heidelberg
Ion Therapy (HIT) Center in Germany for first experiments with helium
ion imaging. Again, TOPAS was used to simulate helium CT data for
predicting and evaluating the characteristics of HeCT images reconstructed
with the DROP-TVS algorithm now in comparison with proton CT images,
see \cite{PFRMSVS2018}. Given the ease and accuracy of results, it
is thus a ``natural'' choice to use TOPAS for forward dose calculations
when testing new IMPT algorithms.

\subsection{Digital Phantoms}

Another critical need is to identify or develop anthropomorphic digital
phantoms necessary for realistic pCT and IMPT algorithm testing. The
general strategy we have followed for most of the algorithms we have
tested has been to demonstrate the algorithmic idea with several phantoms
ranging from simple phantoms often in two dimensions to increasingly
larger and more realistic phantoms. We believe that this approach
prevents the ``masking'' of the principal numerical performance
of new algorithms by difficulties stemming from the complexity of
realistic object data, which may require different adjustments after
the performance of the algorithms has been demonstrated on simpler
datasets. In the following paragraph, we describe the digital head
phantom that we specifically designed to support our pCT algorithmic
developments.

\subsubsection*{Head Phantom}

\begin{figure}[th]
\centerline{\includegraphics[width=3.5in]{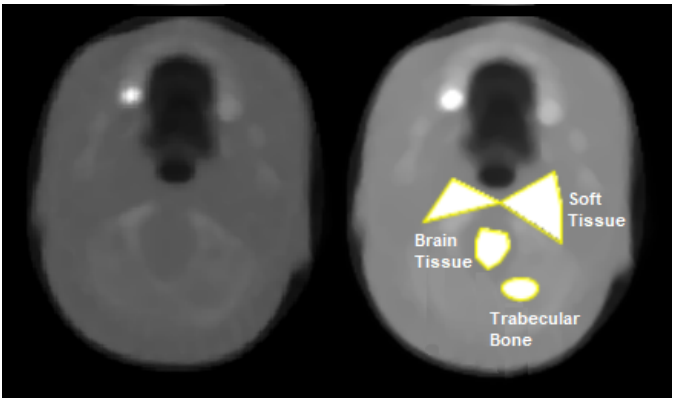}}
\caption[justification = centering]{Representative pCT reconstruction of the slice of the pediatric head
phantom containing regions of interest (left); the regions of interest
are filled in white and labeled by their composition (right). (reproduced
from \cite{BS2019}).}
\label{fig:HN715}
\end{figure}

We created the HIGH\_RES\_HEAD digital head phantom with very high
resolution, \cite{Giacometti2017} for the development and evaluation
of pCT image reconstruction algorithms and IMPT algorithms, but it
can be used for other medical physics purposes as well. The phantom
is now available as a computational DICOM head phantom to the Geant4
user community in the open source Geant4 MC simulation tool package
\cite{GUATELLI2017179}. The phantom represents a head of a 5-year-old
child. It is a very close approximation of the commercial head phantom
(model HN715, CIRS, Norfolk, VA, USA). The HIGH\_RES\_HEAD digital
phantom includes most anatomical details of a human head and is characterized
by a high spatial resolution ($0.18\times0.18\times1.25~mm^{3}$ voxel
size). We used it together with the Geant4 simulation platform of
the pCT system to validate its performance by comparing digital phantom
reconstructions with reconstructions from experimental data \cite{GBPGP17}.
The phantom was also used for testing the parameter space of a novel
superiorization algorithm for pCT reconstruction \cite{BS2019}, described
in Subsection \ref{sec:NTVS} above. Figure \ref{fig:HN715} shows
a representative proton CT reconstruction from that work.

\section{The Loma Linda University algorithm workshops}

We have organized, since 2015, five ``Loma Linda University (LLU)
Workshops on Algorithms and Computational Techniques in Proton Imaging
and Intensity-Modulated Proton Therapy.'' For details about the previous
and planned Annual Loma Linda workshops, see http://ionimaging.org/.
The first three workshops were face-to-face meetings at LLU, Loma
Linda, CA, USA, during the weeks following the annual meetings of
the American Association of Physicists in Medicine (AAPM). Participation
was by invitation only, and we welcomed applied mathematicians, computer
scientists, medical physicists, radiobiologists, and radiation oncologists.
We have organized the workshops in an informal setting: there were
no workshop fees, no parallel sessions, and no strictly timed talks.
The emphasis was on scientific exchange, learning, and discussions.
In all workshop meetings, the participants learned from each other,
and new research projects and collaborations have emerged that are
still on-going. For the 4th and 5th workshops, we also allowed participation
by video-conferencing via Zoom, and the 6th workshop in July 20-22,
2020, was a Zoom-only workshop due to the COVID-19 pandemic.
The 7th workshop was held in August 2-4, 2021 at Loma
Linda via Zoom, see: http://ionimaging.org/llu2021-overview/.

\section{Discussion and Conclusion\label{sec:conclusion}}

When it comes to advanced techniques such as pCT and IMPT, there is
a strong need for input from experts in computer science and mathematics
to support the medical physicist's insight and understanding. With
this in mind, we collaborate on developing advanced reconstruction
algorithms for pCT imaging based on mathematical expertise in iterative
projection methods. With computer science expertise, we have achieved
progress and fast implementation of novel mathematical algorithms
on advanced computing hardware. There is still much more to accomplish
with increasing demands for high-speed and high-performance medical
imaging and verification techniques leading to online-adaptive particle
therapy, the inclusion of biological weighting in treatment planning
optimization or superiorization, and a combination of different image
guidance technologies such as MRI and pCT. Such interdisciplinary
collaboration should stimulate others to follow a similar path and
widen the network of collaborations in the disciplines of medical
physics, mathematical algorithms, and computer science.



\appendix
{Acknowledgments.} We thank the many collaborators on the projects
mentioned here, in particular, Blake Schultze, Paniz Karbasi, Caesar
Ordonez, Christina Sarosiek, Howard Heaton, Jose Ramos-Mendez, Pierluigi
Piersimoni, Robert Jones, Robert Johnson, Stuart Rowland, and Tai
Dou as well as Mark Pankuch and his medical physics and accelerator
support team at the Northwestern Medicine Chicago Proton Center for
help with collecting proton CT data with the Phase II pCT scanner.
The interdisciplinary efforts described here were supported by the
United States-Israel Binational Science Foundation (BSF) (\cite{BSF1},
\cite{BSF2}). The work of Y.C. was supported by the ISF-NSFC joint
research program grant No. 2874/19. All authors contributed equally
to the writing of this paper. All authors read and approved the final
manuscript. We appreciate the anonymous reviewer's
report which helped us improve the paper.




\bibliographystyle{IEEEtran}
\bibliography{new-develops-bib-030721}

%


\begin{IEEEbiography}[{\includegraphics[width=1in,height=1.25in,clip,keepaspectratio]{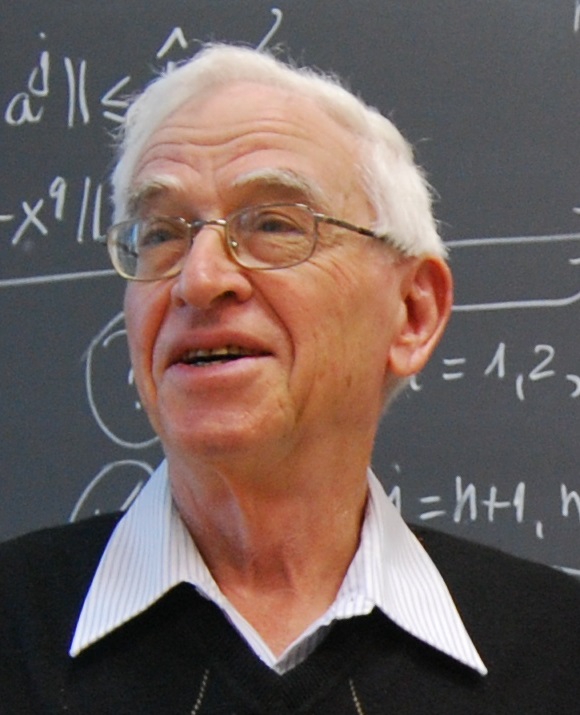}}]{Yair Censor}
received his B.Sc., M.Sc. and D.Sc. degrees in mathematics from
the Technion-Israel Institute of Technology in Haifa in 1967, 1969
and 1975, respectively. From 1977 to 1979, he was Research Assistant
Professor of Computer Science at the State University of New York
(SUNY) at Buffalo, NY, USA, and Research Associate with the Medical
Image Processing Group (MIPG) there.

In 1979, Dr. Censor joined the Department of Mathematics at the University
of Haifa in Israel, where he has been Full Professor since 1989 and
became Professor Emeritus in 2012. He was a visiting professor at
the University of Pennsylvania in Philadelphia; in Link\"{o}ing University
in Sweden; in the Instituto de Matematica Pura e Aplicada in Rio de
Janeiro, Brazil; in the Beijing International Center of Mathematics
at the Peking University in Beijing, China; and in the Graduate Center
of the City University of New York.

Dr. Censor works in computational mathematics, where his interests
include optimization theory, inverse problems, optimization techniques
in image reconstruction from projections, and in intensity-modulated
radiation therapy (IMRT). Dr. Censor has published over 160 research
articles in refereed scientific journals, conference proceedings and
as book chapters. He co-authored with S.A. Zenios the book: Parallel
Optimization: Theory, Algorithms, and Applications, Oxford University
Press, New York, 1997. For this book he received, together with Professor
Zenios, the 1999 ICS (INFORMS Computing Society) Prize for research
excellence in the interface between operations research and computer
science. Dr. Censor served as Associate Editor for the journal IEEE
Transactions on Medical Imaging from 1992 to 2004.
\end{IEEEbiography}

\begin{IEEEbiography}[{\includegraphics[width=1in,height=1.25in,clip,keepaspectratio]{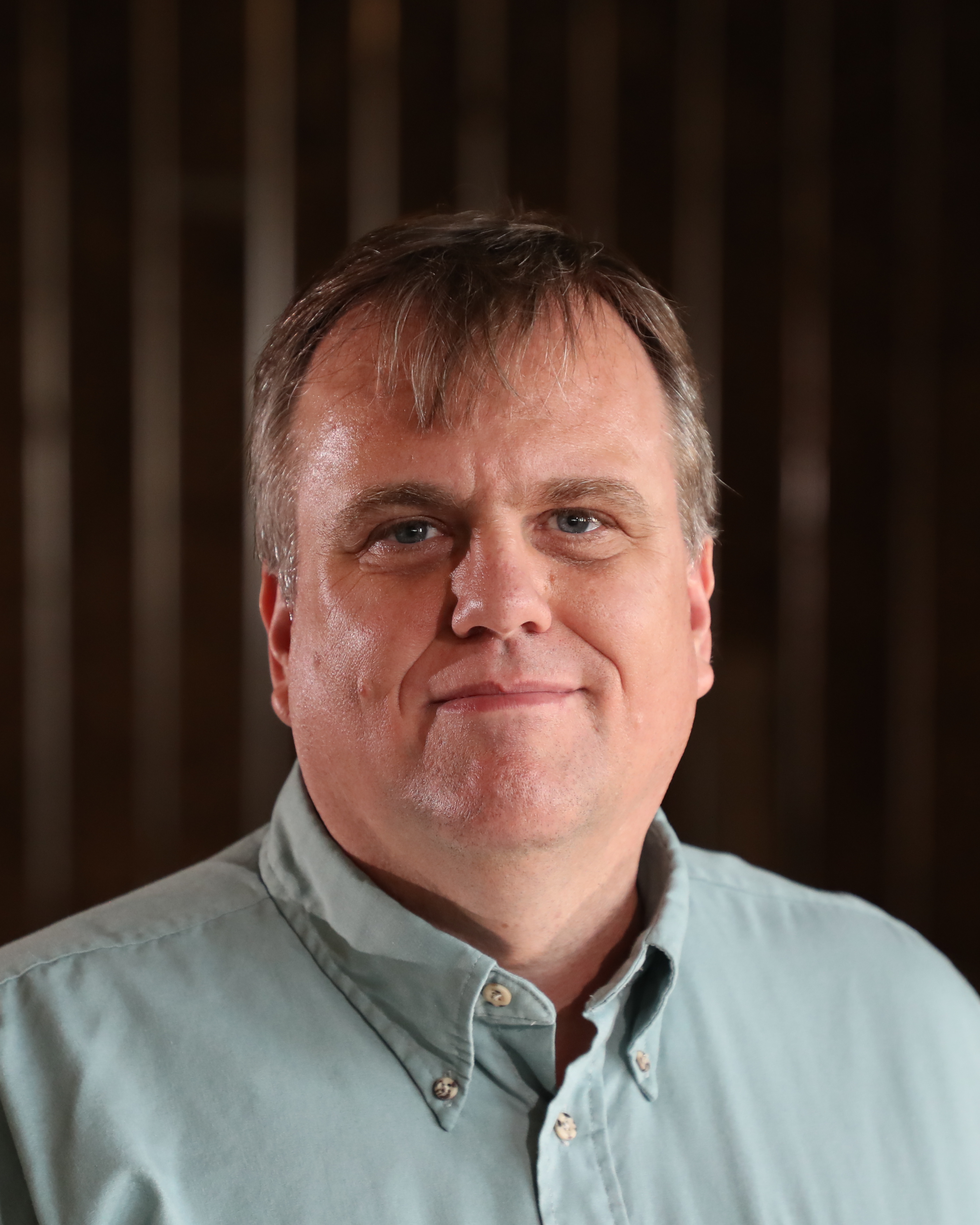}}]{Keith E. Schubert}
received his B.S. degree in general engineering from the University
of Redlands in 1991, his M.S. in electrical engineering from UCLA
in 1992. After working for a few years, he earned his Ph.D. in electrical
and computer engineering from UCSB in 2003 for work on computational
methods of uncertain systems and robustness in estimation and control.

From 1992-1996 he worked for Northrop-Grumman as a design engineer,
and since 1996 he has worked as a consulting engineer for various
companies. In 2000, he took a visiting faculty position at the University
of Redlands in mathematics and computer science for two years, then
was hired as an assistant professor in 2002 at CSU San Bernardino,
where he became an associate professor in 2006, then a full professor
in 2010. At CSUSB he was the lead on starting their first engineering
program and was part of the team that started the Bioinformatics program.
In 2013, he joined Baylor University in Waco, Tx, where he is currently
a full professor in electrical and computer engineering. His research
is in devices and computational methods for robust imaging particularly
for biomedical applications.

Dr. Schubert has several patents and served as Principal Investigator
on the NIH R01 grant to develop proton CT for biomedical applications
from 2011-2016 and as part of the research team for a planning grant
(P20) for the development of ion therapy research in the United States
from 2015-2017. In 2013, Dr. Schubert was one of the team leaders
of a National Geographic research expedition to Cueva de Villa Luz
in Mexico, where he studied patterned biological grows in the sulfuric
acid caves, which was featured in the July 2014 cover article of National
Geographic Magazine.
\end{IEEEbiography}

\begin{IEEEbiography}[{\includegraphics[width=1in,height=1.25in,clip,keepaspectratio]{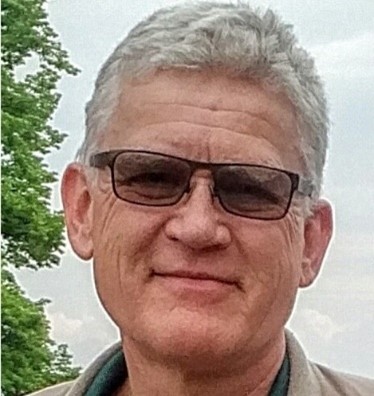}}]{Reinhard W. Schulte}
(M’12) received his diploma degree (M.S. equivalent) in physics
(Diplom Physiker, Dipl. Phys.) from the University of Dortmund, Germany,
in 1978, and his doctorate in medicine (Doctor of Medicine, Dr. med.)
from the University of Cologne, Germany, in 1986. He completed his
residency in radiology and radiation oncology at Hannover Medical
School in 1989.

He was a postdoctoral fellow at the first hospital-based proton therapy
center at Loma Linda University Medical Center (LLUMC) in California
from 1990-1994. In 1994 he accepted a clinical research position at
LLUMC. In 1996, he became an Assistant Clinical Professor in the School
of Medicine and the Department of Radiation Medicine at LLUMC, In
2007, he was promoted to Associate Professor, and since 2013, he has
been a Full Professor at Loma Linda University, now in the Department
of Basic Sciences, Division of Biomedical Engineering Sciences. He
has been successful in developing novel instrumentation and algorithmic
approaches for proton therapy. His current research interests include
the clinical development of proton CT and nanodosimetry to improve
proton and ion therapy. He is an Associate Editor of the journal Medical
Physics and holds several patents.

Prof. Schulte was awarded a habilitation fellowship by the German
Research Foundation (Deutsche Forschungsgemeinschaft, DFG) from 1989-1991.
He is a member of the American Society of Radiation Oncology (ASTRO),
the European Society of Therapeutic Radiation Oncology (ESTRO), the
American Association of Physicists in Medicine (AAPM), the German
Society for Boron Neutron Capture Therapy, and the Radiation Research
Society. He was awarded the ESTRO Calergo Award in 1990. Prof. Schulte
has served as Principal Investigator on several NIH grants, including
an R01 grant to develop proton CT for biomedical applications from
2011-2016 and a planning grant (P20) for the development of ion therapy
research in the United States from 2015-2017.
\end{IEEEbiography}





\end{document}